
\magnification=1200
\vsize=22truecm
\hsize=15truecm \tolerance 1000
\parindent=0pt
\baselineskip = 15pt
\lineskip = 1.5pt
\lineskiplimit = 3pt
\font\twelvebf=cmbx12
\font\twelverm=cmr12

\font\mygoth=eufm10
\def\g{{\hbox{\mygoth g}}}
\def\hg{\hat\g}
\def\ad{{\rm ad}}
\def\U{{\tilde U}}
\def\V{{\tilde V}}
\def\h{{\tilde h}}
\def\H{{\tilde H}}
\def\E{{\hat E}}

\def\NP{{Nucl.\ Phys.\ }}
\def\PL{{Phys.\ Lett.\ }}
\def\IJMP{{Int.\ J.\ Mod.\ Phys.\ }}
\voffset=0pt

\parskip = 1.5ex plus .5ex minus .1ex
{\nopagenumbers
\rightline{KCL-TH-94-8}
\rightline{hep-th/9408092}
\vglue 1truein
\centerline{\twelvebf Conserved charges and soliton solutions}
\centerline{\twelvebf in affine Toda theory}
\bigskip
\centerline{\twelverm Michael Freeman}
\medskip
\centerline{Department of Mathematics}
\centerline{King's College London}
\centerline{Strand, London WC2R 2LS}
\bigskip
\centerline{June 1994}
\vskip .75truein
\centerline{\bf Abstract}
\smallskip
We study the conserved charges of affine Toda field theories by making use
of the conformally invariant extension of these theories.  We compute the
values of all charges for the single soliton solutions, and show that
these are related to eigenvectors of the Cartan matrix of
the finite-dimensional Lie algebra underlying the theory.
\vfil
\eject}

{\twelvebf 1. Introduction}

Affine Toda field theory is a scalar field theory
constructed from an underlying affine Lie algebra~$\hg$,
with many properties of the theory reflecting those of the algebra.
Thus, for example, there are infinitely many conserved charges in
the theory~[1--4], with spins
equal to the exponents of $\hg$,
and by constructing S-matrix elements for
the theory it has been observed that the values of these
charges form eigenvectors of the Cartan matrix of
the finite-dimensional Lie algebra $\g$ associated to $\hg$~[5--10].
The allowed three-point couplings were also seen to be related to
$\g$~[5,6,11--13],
and in~[14] Dorey gave a characterization of these couplings
that explained many of the above results.
In a separate development, an explanation was given of the Lie-algebraic
nature of the classical mass spectrum and couplings~[15,16] starting from
the Lagrangian of the theory.

It was subsequently proposed by Hollowood~[17] to study the theory with
an imaginary value of the coupling constant, for which there are
infinitely many degenerate vacuum states with soliton solutions
interpolating between them.  Such solutions were constructed first
for the $A_n^{(1)}$ theories, and although they are not real
they nevertheless have real energy.  Further soliton solutions were found
for other Lie algebras in~[18-24], with the most general solution
being given in~[18,19], and the energy and momentum were
found to be real in these cases also.  The masses of the single
soliton solutions were seen to form the {\it left} Perron-Frobenius
eigenvector of the Cartan matrix
of $\g$~[18], while the masses of the fundamental excitations
formed the {\it right} Perron-Frobenius eigenvector.
These results provide evidence for a conjectured duality between affine
Toda field theories, in which
the solitons of one theory correspond
to the fundamental excitations of another.

In this paper we extend the calculation of Olive, Turok and Underwood~[18]
by computing the
classical values of all conserved charges for the single soliton solutions.
We shall see that these are all real, apart from a possible overall
phase factor, and that the values of the charge of spin-$N$ form the
left eigenvector of the Cartan matrix of $\g$ with eigenvalue
$2(1 - \cos\pi N/h)$, where $h$ is the Coxeter number of $\g$.
This provides further evidence for the duality of affine
Toda theories.

The conserved charges are also the subject of interesting work by
Niedermaier~[25,26], in which expressions for the charges are obtained by
the inverse scattering method for the case $\hg=A_n^{(1)}$.  It would
be worthwhile to extend this approach to arbitrary affine algebras and to
relate it to that given here.

{\twelvebf 2. Untwisted affine Lie algebras and the affine Toda equation}

We first review those concepts from the theory of untwisted affine Lie
algebras that will be needed in this paper~[27].
\medskip
{\bf 2.1 Affine Lie algebras}

Let $\g$ be a finite-dimensional simple Lie algebra  of rank r,
with simple roots $\alpha_1, \ldots, \alpha_r$ and Cartan matrix
$K_{ij} = 2 \alpha_i\cdot\alpha_j/\alpha_j^2$.
The so-called extended Cartan matrix is constructed similarly
from an extended set of roots, obtained by adjoining
to the set of simple roots an
extra root $-\psi$, where $\psi$ is the highest root of $\g$.
This extra root will be denoted $\alpha_0$, and the extended Cartan
matrix will again be written as $K_{ij}$ but with $i$ and
$j$ taking values from 0 to $r$.  This matrix will be the Cartan matrix of
the untwisted affine Lie algebra $\hat \g$ corresponding to $\g$.
In the Chevalley basis $\hg$ is generated by elements $e_i, f_i, h_i$,
for $i=0,\ldots, r$, satisfying the commutation relations
$$
\eqalign{
[h_i,h_j]&=0, \cr
[h_i,e_j]&=K_{ji}e_j, \cr
[h_i,f_j]&=-K_{ji}f_j, \cr
[e_i,f_j]&=\delta_{ij}h_i
}\eqno(2.1)
$$
and subject to the Serre relations
$$
(\ad\> e_i)^{1-K_{ji}}e_j = 0 \eqno(2.2)
$$
and
$$
(\ad\> f_i)^{1-K_{ji}}f_j = 0. \eqno(2.3)
$$
With respect to this basis there is a natural grading of the
algebra, called the principal grading, such that the elements $e_i$ have
grade 1 and the $f_i$ grade $-1$.\footnote\dag{It is often useful to
add to this algebra an extra element called the derivation, the
adjoint action of which produces
this grading, but we shall not do this here.}
The elements of grade $>0$
generate a subalgebra $\hat n_+$ and those of grade $< 0$
a subalgebra $\hat n_-$.  We shall assume that it is
legitimate to exponentiate these subalgebras where necessary, thereby obtaining
groups $\hat N_\pm$, and we shall write $\hat T$ for the group obtained by
exponentiating elements of grade zero.

The fact that the highest root of $\g$ has an expansion in terms of
simple roots implies the existence of a relation
$\sum_0^r n_i \alpha_i = 0$,
where the $n_i$ are positive integers and $n_0=1$.
{}From this it follows that the Cartan matrix of $\hg$ has a left-eigenvector
with eigenvalue zero,
$$
\sum_{i=0}^r n_i K_{ij} = 0. \eqno(2.4)
$$
By considering the Lie algebra $\check\g$ whose roots
$\check\alpha=2\alpha/\alpha^2$ are the co-roots of $\g$
it follows similarly that
$$
\sum_{j=0}^r K_{ij} m_j = 0, \eqno(2.5)
$$
with $m_j$ positive integers and $m_0 = 1$.
The Coxeter number $h$
and dual Coxeter number $\tilde h$ of $\g$ are defined by
$h=\sum_{i=0}^r n_i$ and $\tilde h = \sum_{i=0}^r m_i$.

A consequence of eqn~(2.1) is that $\hg$ contains a central element
$x = \sum_0^r m_i h_i$, with $[x,e_i]=[x,f_i]=0$.
Let $\E_1$ and $\E_{-1}$ be linear
combinations of the $e_i$ and $f_i$ respectively such that
$$
[\E_1, \E_{-1}] = x; \eqno(2.6)
$$
this condition does not determine $\E_{\pm 1}$ uniquely, but
a convenient choice is $\E_1=\sum_i \sqrt{m_i} e_i$ and
$\E_{-1} = \sum_i \sqrt{m_i} f_i$.  Most of our results, however,
will be independent of any particular choice.
Consider now the set of non-trivial
elements of $\hg$ that
commute with $\E_{\pm 1}$ modulo the central element $x$. Each such
element can be chosen to have definite principal grade, and
it can be shown~[27] that an element
$\E_N$ of grade $N$
exists precisely when $N$ is equal to an exponent of $\g$ modulo the Coxeter
number.  Furthermore, with a suitable choice of normalisation we have
$$
[\E_N, \E_M] = x N \delta_{N+M,0}, \eqno(2.7)
$$
so the elements $\E_N$ and $x$ span an infinite-dimensional Heisenberg
subalgebra of $\hg$.
\medskip
{\bf 2.2 The affine Toda equations}

Affine Toda field theory is formulated in terms
of a field $\phi$ taking values in the grade zero subalgebra $\hg_0$ of
$\hg$, spanned by $h_0, \ldots, h_r$.
The equation of motion for $\phi$ is
$$
\beta\partial_\mu\partial^\mu\phi +
4 m^2 [e^{\beta\phi}\E_1e^{-\beta\phi},\E_{-1}] = 0, \eqno(2.8)
$$
which is independent of any particular choice that is made for
$\E_{\pm 1}$ satisfying eqn~(2.6).  It should be remarked that this equation
differs slightly from the conventional form of the affine Toda equation,
but this can be recovered from eqn~(2.8) by expanding $\phi$ in the form
$\phi=\sum_1^r \phi_i h_i + \phi_x x$. The equations for $\phi_i$ are
$$
\beta\partial_\mu\partial^\mu\phi_i + 4 m^2 m_i
\left(e^{\beta\sum_{j=1}^r K_{ij}\phi_j} - e^{\beta\sum_{j=1}^r K_{0j}\phi_j}
\right)=0, \eqno(2.9)
$$
which are just the usual affine Toda equations, while $\phi_x$ is determined
non-locally in terms of the $\phi_i$ by the equation
$$
\beta\partial_\mu\partial^\mu\phi_x + 4 m^2 e^{\beta \sum_{j=1}^r K_{0j}\phi_j}
=0. \eqno(2.10)
$$
Eqns~(2.9) and~(2.10) are the conformal affine Toda equations~[28], with the
field corresponding to the derivation set to zero.

It will be useful to introduce
light-cone coordinates $x^{\pm}=x\pm t$, with respect to which the
equation of motion~(2.8) becomes
$$
-\beta\partial_+\partial_-\phi +
m^2 [e^{\beta\phi}\E_1e^{-\beta\phi},\E_{-1}] = 0. \eqno(2.11)
$$
This can be written as a zero curvature condition,
$$
[\partial_+ + A_+, \partial_- + A_-] = 0, \eqno(2.12)
$$
with
$$
A_+ = m e^{\beta\phi}\E_1 e^{-\beta\phi} \eqno(2.13)
$$
and
$$
A_- = m \E_{-1} - \beta\partial_-\phi. \eqno(2.14)
$$
The standard procedure used to demonstrate the integrability of this system
is to make a gauge transformation $A_\mu \rightarrow
A_\mu^\omega = \omega^{-1}\partial_\mu\omega + \omega^{-1}A_\mu\omega$
such that $A_-^\omega$ lies entirely in the subspace of $\hg$ spanned
by the $\E_N$ with $N\ge -1$.
This is possible with $\omega$
having the form $\omega=\exp\{\omega_1 + \omega_2 + \ldots\}$, where
$\omega_i$ is a local function of $\partial_-\phi$ with grade $i$.
The equation of motion then implies that $A_+^\omega$ lies in the span
of the $\E_N$ with $N\ge 2$, from which it follows that
$A_-^\omega$ and $A_+^\omega$ commute.  In this gauge the equation of
motion takes the form
$$
\partial_+A_-^\omega = \partial_-A_+^\omega, \eqno(2.15)
$$
which implies immediately the existence of an infinite number
of conserved currents.  These can conveniently be expressed by taking
the commutator of eqn~(2.15) with $\E_{-N}$, giving an infinite set of
conserved
currents taking values in the centre of $\hg$,
$$
\partial_+[A_-^\omega, \E_{-N}] = \partial_-[A_+^\omega,\E_{-N}]. \eqno(2.16)
$$

Unfortunately this procedure for constructing conserved charges suffers
from a problem in the present context.  The gauge transformation
$A_- \rightarrow A_-^\omega$ is not uniquely determined, since there is a
freedom to multiply $\omega$ on the right by the exponential of any linear
combination of elements of the Heisenberg subalgebra with grade $\ge 2$.
This changes the conserved currents by a total derivative, and although
this would not normally have any effect on the conserved charges it does here.
The reason for this is that even though the fields $\phi_i$ and
their derivatives can be taken to vanish at infinity, the same is not
true for $\phi_x$, because this field must satisfy eqn~(2.10).
For the vacuum solution $\phi_i=0$, for example, we find
$\phi_x = m^2 x^+ x^- / \beta$,
and in fact the boundary terms coming from integrating
derivatives of $\phi_x$ will be crucial in obtaining the conserved charges
of soliton solutions.

There is, however, another form for the conserved currents, which does
not suffer from these disadvantages.  This alternative form was first given
by Wilson~[1] for the unextended affine Toda theory~(2.9), and in that
context the relation to the construction outlined above is somewhat obscure.
In the extended theory we are considering, however, the connection
between these two approaches follows immediately from an attempt to
construct commuting flows for the equation~(2.8).
\medskip
{\bf 2.3 Commuting flows in extended affine Toda theory}

In the construction of commuting flows it is convenient to
regard $x^-$ as the spatial coordinate and to consider $x^+$ as
the time direction.  We introduce an infinite collection
of additional time variables $t_{-N}$ and define the evolution  of $\phi$ with
respect to these by the zero curvature conditions
$$
\left[{\partial\over\partial x^-} + A_-, {\partial\over\partial t_{-N}}
+ A_{-N}\right]=0\eqno(2.17)
$$
for some suitable $A_{-N}$.
A natural choice would be to take $A_{-N}$ equal to
$m^{N} \sum_{n\le 0}(\omega \E_{-N} \omega^{-1})_n$,
with $\omega$ being the group element introduced earlier and
with $(\omega \E_{-N}\omega^{-1})_n$ denoting the terms of
grade $n$ in $(\omega \E_{-N}\omega^{-1})$.  This generalises $A_-$, in that
$A_{-1} = A_-$, and in the non-extended theory the analogue of this
construction leads to an infinite set of commuting flows.
In the extended theory, however, the evolution of the additional
field $\phi_x$ depends explicitly on the choice of $\omega$, and
the evolution operators $\partial_{-N}$ do not commute either
amongst themselves or
with $\partial_+$ when acting on $\phi_x$.  In this case a
better choice is to take instead
$$
A_{-N} = - m^{N} \sum_{n > 0}(\omega \E_{-N} \omega^{-1})_n,\eqno(2.18)
$$
which leads to the same evolution equations in the non-extended theory
but does not suffer from the above difficulties.

In order for the evolution equation~(2.17) to be consistent, it is
essential that $\partial_{-N}\phi$
as determined by it should have grade zero.  This follows from
the fact that $A_{-N}$ can be written in two distinct ways,
$$
\eqalignno{
A_{-N} &= - m^{N} \sum_{n > 0}(\omega \E_{-N} \omega^{-1})_n&(2.19) \cr
&= - m^{N} (\omega \E_{-N}\omega^{-1}) +
m^{N}\sum_{n\le 0}(\omega \E_{-N} \omega^{-1})_n.&(2.20)
}
$$
{}From these we obtain the following two expressions for
$\partial_{-N}\partial_-\phi$,
$$
\eqalignno{
\beta{\partial\over\partial t_{-N}}\partial_-\phi &=
m^{N+1} [\E_{-1},(\omega \E_{-N} \omega^{-1})_1] &(2.21)\cr
&=m^{N} [A_-^\omega, \E_{-N}] -
m^{N}\partial_-(\omega \E_{-N}\omega^{-1})_0,&(2.22)
}
$$
and from the first of these we see that, as claimed above, the evolution
$\partial_{-N}$
does not depend on the particular choice of $\omega$ that we make,
because the quantity $[\E_{-1},(\omega \E_{-N} \omega^{-1})_1]$ is unchanged
under the group of transformations that leaves
the form of $A_-^\omega$ invariant,
$$
\eqalign{
\left[\E_{-1},\left(\omega e^{b_N \E_N}\right) \E_{-N}
\left( e^{-b_N \E_N}\omega^{-1}\right)\right] &=
[\E_{-1},\omega ( \E_{-N} + N b_N x ) \omega^{-1})] \cr
&= [\E_{-1},(\omega \E_{-N} \omega^{-1})].
}\eqno(2.23)
$$
To verify the commutativity of the flows, we consider the action of the
evolution operators on $A_-$.  Thus, for example,
$[\partial_{-M}, \partial_{-N}]A_-$ is given by
$$[\partial_{-M}, \partial_{-N}]A_- =
\bigl[\partial_- + A_-,
\partial_{-M} A_{-N} -\partial_{-N}A_{-M} +[A_{-M},A_{-N}]\bigr],\eqno(2.24)
$$
and this can be seen to vanish when we insert the explicit expressions
for $A_{-N}$ and $A_{-M}$ from eqn~(2.19) and make use of
the fact that $A_{-N}^\omega$ and $A_{-M}^\omega$ are
forced, as a consequence of eqn~(2.17), to lie in the Heisenberg subalgebra
of $\hg$.  The vanishing of $[\partial_+,\partial_{-N}]$ is essentially
similar except that in this case we need to evaluate
$[\partial_{-N}\phi, A_+]$, while from eqns~(2.21, 2.22) we know only
$\partial_{-N}\partial_-\phi$.  From eqn~(2.22), however, we see that
$$
\partial_-\left\{\beta{\partial\over\partial t_{-N}}\phi +
m^N(\omega \E_{-N}\omega^{-1})_0\right\}\eqno(2.25)
$$
will vanish in any commutator, so that within a commutator it will be
consistent to set $\beta\partial_{-N}\phi$ equal to
$- m^N(\omega \E_{-N}\omega^{-1})_0$.  Provided we make this choice,
the flows $\partial_+$ and $\partial_{-N}$ will commute.

Comparing the two expressions~(2.21) and (2.22)
we see that\footnote\dag{We note that eqn~(2.26) could have been obtained
immediately, without introducing the time variable $t_{-N}$, by
evaluating directly the expression
$\partial_-(\omega \E_{-N}\omega^{-1})_0$ and making use of the fact that
$A_-^\omega$ lies in the Heisenberg subalgebra.}
$$
m [\E_{-1},(\omega \E_{-N} \omega^{-1})_1] = [A_-^\omega, \E_{-N}]
- \partial_-(\omega \E_{-N}\omega^{-1})_0.\eqno(2.26)
$$
Thus
$m [\E_{-1},(\omega \E_{-N} \omega^{-1})_1]$ is a conserved density, since
it differs from  $[A_-^\omega, \E_{-N}]$
by a derivative term.
This is the form for the conserved densities given by Wilson~[1].
It should be remarked that only the terms proportional to the
central element $x$ in eqn~(2.26) are of interest, because
$[A_-^\omega, \E_{-N}]$ lies in the centre of $\hg$ and so all components
of $[\E_{-1},(\omega \E_{-N} \omega^{-1})_1]$ other than that proportional
to $x$ are given by derivatives of local fields.  The corresponding
charges therefore vanish.

It is interesting to note that the field $\phi_x$ does not enter into
the expression $[\E_{-1},(\omega \E_{-N} \omega^{-1})_1]$ for the
conserved densities~[4].  To see this we exploit the fact that these quantities
are independent of $\omega$ to make the following convenient choice.
We first
find $\omega_0$ depending only on $\phi_1, \ldots, \phi_r$ such that
$A_-^{\omega_0}$ takes the form
$$
A_-^{\omega_0} = m \E_{-1} -\beta\partial_-\phi_x x +
\sum_{N>0} a_N \E_N;\eqno(2.27)
$$
$\omega_0$ is just the gauge transformation that would be used in the
non-conformal affine Toda field theory to transform $A_-$ and $A_+$ into an
abelian subalgebra of $\g$.  We can now make a further transformation by
$\exp{(\phi_x \E_1)}$ to remove the central term from eqn~(2.27), but
the argument given above implies that this has no
effect on $[\E_{-1},(\omega \E_{-N} \omega^{-1})_1]$ and so the conserved
densities are indeed independent of $\phi_x$.

In addition to the conserved densities
$[\E_{-1},(\omega \E_{-N} \omega^{-1})_1]$ of negative spin, there
is another set of conserved densities with positive
spin.  These are obtained by first making a gauge transformation on
the potentials $A_+$ and $A_-$ to bring them into the form
$$
A_+ = m \E_{1} + \beta\phi\eqno(2.28)
$$
and
$$
A_- = m e^{-\beta\phi}\E_{-1}e^{\beta\phi}.\eqno(2.29)
$$
We then make a further gauge transformation $A_\mu\rightarrow A_\mu^{\omega'}$
such that $A_+^{\omega'}$ lies in the subspace of $\hg$ spanned by the
$\E_N$ with $N\le 1$.  Conserved densities of positive spin can then be
constructed by a procedure entirely analogous to that given above.
{}From here on we shall treat only the negative
spin densities, but equivalent results will hold for those of positive spin.
\bigskip
{\twelvebf 3. The conserved charges of soliton solutions}

In order to calculate the values of the conserved charges for soliton
solutions it is necessary to obtain explicit expressions
for $\partial_{-N}\phi$; the conserved charges will then follow from
eqns~(2.21) and~(2.22).
\medskip
{\bf 3.1 The Leznov-Saveliev solution and the conserved charges}

We shall use the method of Leznov and Saveliev~[29] to solve the
Toda equations.  This is closely related to the approach of Date
et al~[30]; the relation between these methods is explained in reference~[31].
The starting point is the zero curvature condition
$$
[\partial_- + A_-,\partial_+ + A_+] = 0,\eqno(3.1)
$$
which implies the existence of some monodromy matrix
$T$ such that
$$
\eqalign{
A_+ &= T^{-1}\partial_+ T \cr
A_- &= T^{-1}\partial_- T \cr
}\eqno(3.2)
$$

Following Leznov and Saveliev we now consider two different Borel
decompositions of $T$,
$$
T = U H V^{-1},\qquad U\in \hat N_+, \;H\in \hat T, \;V\in \hat N_-,\eqno(3.3)
$$
and
$$
T= \V \H \U^{-1},\qquad \V\in N_-,\; \H\in T,\;\U\in N_+.\eqno(3.4)
$$
We shall write $H=e^h$ and $\H=e^\h$.
{}From the equation for $A_+$ we obtain
$$
\eqalignno{
\partial_+\V&=0&(3.5) \cr
\partial_+\h&=0&(3.6) \cr
\U\partial_+\U^{-1} &= m e^{\beta\phi}\E_1 e^{-\beta\phi}&(3.7) \cr
H^{-1}U^{-1}\partial_+UH &= m e^{\beta\phi} \E_1 e^{-\beta\phi},&(3.8)
}
$$
and from $A_-$ we have
$$
\eqalignno{
\partial_-U&=0 &(3.9)\cr
\partial_-h + \beta\partial_-\phi&=0 &(3.10)\cr
\H^{-1}\V^{-1}\partial_-\V\H &= m \E_{-1} &(3.11)\cr
\U\partial_-\U^{-1} + \U(m \E_{-1} + \partial_-\h)\U^{-1} &=
m \E_{-1} - \beta\partial_-\phi.&(3.12)
}
$$
A convenient way to obtain solutions to the affine Toda equations
is to consider the group element
$$
\eqalign{
g&=e^{-\beta\phi}V^{-1}\U \cr
&= e^{-\beta\phi} H^{-1}U^{-1} \V\H.
}\eqno(3.13)
$$
{}From eqns~(3.5--3.12) it follows that
$$
\eqalign{
\partial_+g g^{-1}&=-m\E_1 -\partial_+(\beta\phi+h) \cr
g^{-1}\partial_-g &= m\E_{-1} + \partial_-\h.
}\eqno(3.14)
$$
In addition we have $\partial_-(\partial_+g g^{-1})=
\partial_+(g^{-1}\partial_-g)=0$, so that $g$ can be thought of as a
constrained WZW field~[32].  The general solution to the affine
Toda equations is obtained by solving~(3.14) to express $g$ in terms
of the free fields $\h$ and $h + \beta\phi$ and then recovering
$\phi$ by taking the matrix elements of $g$ between general highest weight
states.  The soliton solutions are precisely those with $\h=0$
and $h=-\beta\phi$~[18], in which case
$$
g(x^+,x^-)=e^{-m\E_1 x^+} g(0)e^{m\E_{-1}x^-},\eqno(3.15)
$$
where the group element $g(0)$ is a constant of integration.
Then, if $|\Lambda\rangle$ is any highest weight state, soliton
solutions are given by
$$
\langle\Lambda| e^{-\beta\phi}|\Lambda\rangle =
\langle\Lambda|e^{-m\E_1 x^+} g(0)
e^{m\E_{-1}x^-}|\Lambda\rangle.\eqno(3.16)
$$
Furthermore, when $\h$ and $h+\beta\phi$ vanish, a candidate for the
group element $\omega$ is $\U$, since from eqns~(3.12) and~(3.7) we see that
$$
A_-^\U = m \E_{-1},\qquad A_+^\U = 0.\eqno(3.17)
$$

In order to find the conserved charges of soliton solutions we need to
evaluate the term in the current $[\E_{-1},(\omega \E_{-N}\omega^{-1})_1]$
that is proportional to the central element $x$.  To do this we can
take the expectation of this current in the highest weight state
$|\Lambda_0\rangle$ of the vacuum representation
of $\hg$ at level one.  From eqns~(2.26) and~(3.17), with $\omega=\U$, this is
given by
$$
m \langle\Lambda_0|[\E_{-1},(\U \E_{-N}\U^{-1})_1]|\Lambda_0\rangle
=-\partial_- \langle\Lambda_0|(\U \E_{-N}\U^{-1})|\Lambda_0\rangle.\eqno(3.18)
$$
The group element $\U^{-1}$ acts as the identity on highest weight states,
and from the definition of $g$ in eqn~(3.13) we see that
$$
\langle\Lambda_0|\U \E_{-N}|\Lambda_0\rangle =
{\langle\Lambda_0|g \E_{-N}|\Lambda_0\rangle \over
\langle\Lambda_0|g|\Lambda_0\rangle}.\eqno(3.19)
$$
The general expression for the conserved charge is therefore
$$
\eqalign{
Q_{-N} & \equiv \int_{-\infty}^{\infty} dx^{-}\,
m^N \langle\Lambda_0|[\E_{-1},(\omega \E_{-N}\omega^{-1})_1]|\Lambda_0\rangle
\cr
&= \left[m^N
{\langle\Lambda_0|g \E_{-N}|\Lambda_0\rangle \over
\langle\Lambda_0|g|\Lambda_0\rangle}
\right]_{x^-=-\infty}^\infty.
}\eqno(3.20)
$$
Just as for the energy-momentum tensor, the only contribution for soliton
solutions is a surface term.
\medskip
{\bf 3.2 Conserved charges and the Cartan matrix}

For the single soliton solutions, eqn~(3.20) is particularly simple to
evaluate.  These solutions are characterised by $g(0)$ having the form
$g(0)=\exp\{Q \hat F(\alpha,\rho)\}$~[18], where  $Q$ is a constant
and $\hat F(\alpha,\rho)$ is an element of $\hg$ that diagonalises the
adjoint action of the Heisenberg subalgebra,
$$
[\E_N, \hat F(\alpha,\rho)] =
(\alpha,E_\nu) \rho^N \hat F(\alpha,\rho).\eqno(3.21)
$$
Here $\rho$ is a constant, $\alpha$ is a root of $\g$, and
$E_\nu$ is the element of $\g$ of which $\E_N$ is the abelianization;
$\nu$, which is equal to $N$ mod $h$, is an exponent of $\g$.  It can be
shown that the $E_\nu$ span a Cartan subalgebra of $\g$~[33], and
$(\alpha,E_\nu)$ is the natural pairing between roots and
elements of this Cartan subalgebra.
These results follow from the isomorphism of
the affine Lie algebra $\hg$ to the central extension of the loop algebra
of $\g$~[27].  Using these results the single soliton solution can be written
in the form
$$
\langle\Lambda| e^{-\beta\phi}|\Lambda\rangle =
\langle\Lambda|e^{-x m^2 x^+x^-}
\exp\left\{Q(x^+,x^-) \hat F(\alpha,\rho)\right\}
|\Lambda\rangle,\eqno(3.22)
$$
where
$$
Q(x^+,x^-) = Q e^{-mx^-(\alpha,E_{h-1})\rho^{-1}
-mx^+(\alpha,E_{1})\rho}.\eqno(3.23)
$$
For static solitons, which is all that we shall consider here,
the parameter $\rho$ must satisfy the condition
$$
(\alpha,E_1)\rho = (\alpha,E_{h-1})\rho^{-1};\eqno(3.24)
$$
$Q(x^+,x^-)$ will then be independent of $t$, and the only time-dependence
in $\phi$ will be in the field $\phi_x$.

To evaluate explicitly the values of the charges for these static solutions
we write
$$
\eqalign{
\langle\Lambda_0|g(x^+,x^-) \E_{-N}|\Lambda_0\rangle &=
-(\alpha, E_{h-\nu}) \rho^{-N} Q(x^+,x^-)\times \cr
&\qquad\langle\Lambda_0|e^{-x m^2 x^+x^-}\hat F(\alpha,\rho)
\exp\left\{Q(x^+,x^-) \hat F(\alpha,\rho)\right\}
|\Lambda_0\rangle,
}\eqno(3.25)
$$
using the commutation relations of $\hat F$ with the Heisenberg
subalgebra.  The value of the charge $Q_{-N}$ is then
$$
\eqalign{
Q_{-N} &= -(\alpha, E_{h-\nu})\left({m\over\rho}\right)^{N} \cr
&\qquad\times\left[Q(x^+,x^-)
{\langle\Lambda_0|\hat F(\alpha,\rho)
\exp\left\{Q(x^+,x^-) \hat F(\alpha,\rho)\right\}
|\Lambda_0\rangle
\over
\langle\Lambda_0|
\exp\left\{Q(x^+,x^-) \hat F(\alpha,\rho)\right\}
|\Lambda_0\rangle}\right]_{x^- = -\infty}^\infty
}\eqno(3.26)
$$
Since the highest non-vanishing power of $\hat F(\alpha,\rho)$
in a representation of level $x$ is $2x/\alpha^2$~[19], it is easy to
see by expanding the exponentials in this expression as a power series that
$$
Q_{-N} = \mp {2\over\alpha^2} (\alpha, E_{h-\nu})
\left({m\over\rho}\right)^{N}\eqno(3.27)
$$
for a soliton at rest,
where the sign depends on whether $(\alpha,E_{h-1})\rho^{-1}$ is
positive or negative.

Alternatively we could use eqn~(3.26) to derive
the relation
$$
Q_{-N} = {(\alpha, E_{h-\nu})\over(\alpha, E_{h-1})}
\left({m\over\rho}\right)^{N-1} Q_{-1}\eqno(3.28)
$$
and then find the values of $Q_{-N}$ by using the fact that the
values of the charge $Q_{-1}$ are known~[18]
to be proportional to the entries in the left Perron-Frobenius eigenvector
of the Cartan matrix of $\g$.

To complete the evaluation of the conserved charges we need to know the
quantities $(\alpha,E_{\nu})$, which can be calculated~[15,16]
using the fact that the elements $E_\nu$ are eigenvectors of a
Coxeter element $\sigma$ of the Weyl group $W$ of $\g$~[33].  Let us
review briefly this procedure, which relates eigenvectors of the Coxeter
element to those of the Cartan matrix~[34,35].  The easiest way to do this is
to make use of the isomorphism between the Cartan subalgebra $H$ of $\g$
and its dual $H^*$, spanned by the roots of $\g$~[16].  Thus, to each
element $E_\nu$ there corresponds an element $q_\nu$ such that
$$
(E_\nu,h) = (q_\nu,h) \qquad\forall\; h\in H,\eqno(3.29)
$$
where $(E_\nu,h)$ denotes the natural inner product on $H$ and
$(q_\nu,h)$ the pairing between elements of $H^*$ and $H$.  By means of
this isomorphism $H^*$ inherits an inner product from that on $H$, and
using this we can write the relation between
$E_\nu$ and $q_\nu$ as
$$
(\alpha, E_\nu) = (\alpha, q_\nu).\eqno(3.30)
$$
We shall therefore look for the eigenvectors $q_\nu$ of $\sigma$
acting on the root space.
We label the points of the Dynkin diagram of $\g$ alternately black and
white, and write $\sigma = \sigma_B\sigma_W$,
where $\sigma_B$ (respectively $\sigma_W$) is a product of Weyl
reflections in all of the black (white)
simple roots.  Writing the Cartan matrix of $\g$ as $K = 2 - C$, so that
$C$ is the incidence matrix of the Dynkin diagram of $\g$ and takes the
block form
$$
C = \left(\matrix{0&C_{wb}\cr
                  C_{bw}&0\cr}\right),\eqno(3.31)
$$
it is
elementary to show that
$$
\eqalign{
\sigma_W(\alpha_w) &= - \alpha_w \cr
\sigma_B(\alpha_w) &= \alpha_w + \sum_{b\in B} C_{wb}\alpha_b
}\eqno(3.32)
$$
and
$$
\eqalign{
\sigma_B(\alpha_b) &= - \alpha_b \cr
\sigma_W(\alpha_b) &= \alpha_b + \sum_{w\in W} C_{bw}\alpha_w.
}\eqno(3.33)
$$
Suppose now that $(x_w^\nu, x_b^\nu)$ is a left eigenvector of $C$ with
eigenvalue $2\cos{\pi\nu/h}$,
$$
\sum_b x_b^\nu C_{bw} = 2\cos({\pi \nu/h}) x_w^\nu,
\qquad \sum_w x_w^\nu C_{wb}= 2\cos({\pi \nu/h}) x_b^\nu.\eqno(3.34)
$$
Then $(x_w^\nu, -x_b^\nu)$ is also an eigenvector of $C$, with
eigenvalue $2\cos{\pi(h-\nu)/h}$, so we can take
$(x_w^{h-\nu}, x_b^{h-\nu})$
to be proportional to $(x_w^\nu, -x_b^\nu)$.  The constant of proportionality
can always be chosen to be $1$, except for the case $\nu=h/2$ when it can
be either $\pm 1$.  For simplicity we shall not treat the case
$\nu=h/2$ here, although the analysis is similar.  The right eigenvectors of
$C$ are given by $(y_w^\nu,y_b^\nu)=(x_w^\nu \alpha_w^2,
x_b^\nu \alpha_b^2)$, and we will choose the normalisation
$$
\sum_w x_w^\nu y_w^\mu + \sum_b x_b^\nu y_b^\mu=h\delta_{\nu,\mu}.\eqno(3.35)
$$

The eigenvectors of the Coxeter element $\sigma$ can now be written
in terms of those of the Cartan matrix.  The
eigenvector with eigenvalue $\exp(2\pi i \nu/h)$
can be chosen to be
$$
\exp\left(-\pi i \nu\over 2h\right)\sum_w x_w^\nu \alpha_w +
\exp\left(\pi i \nu\over 2h\right) \sum_b x_b^\nu \alpha_b,\eqno(3.36)
$$
and given the normalization $(E_\nu, E_\mu)=h\delta_{\nu+\mu,h}$ it is easy
to see that the correctly normalized $q_\nu$ are given in terms of these
eigenvectors by
$$
q_\nu={e^{\pi i/4}\over \sin \pi\nu/h}
\left\{
\exp\left(-\pi i \nu\over 2h\right)\sum_w x_w^\nu \alpha_w +
\exp\left(\pi i \nu\over 2h\right) \sum_b x_b^\nu \alpha_b
\right\},\eqno(3.37)
$$
from which we obtain
$$
(\alpha_w,q_\nu)=\exp(\pi i\nu/2h - \pi i/4)y_w^\nu\eqno(3.38)
$$
and
$$
(\alpha_b,q_\nu)=\exp(-\pi i\nu/2h +3\pi i/4)y_b^\nu.\eqno(3.39)
$$

We now have all the information we need to compute the charges of the
solitons.  There are $r$ species of solitons, given by taking
$\alpha$ equal to $\alpha_w$ and $-\alpha_b$ in $\hat F(\alpha,\rho)$~[18].
Let us consider the case $\alpha=\alpha_w$ first.  We have
$$
(\alpha_w,q_1)=\exp(\pi i/2h - \pi i/4) y_w^1\eqno(3.40)
$$
and
$$
(\alpha_w,q_{h-1})=\exp(-\pi i/2h + \pi i/4) y_w^1,\eqno(3.41)
$$
from which it follows
$$
\rho =\mp \exp(\pi i/4-\pi i/2h),\eqno(3.42)
$$
where the minus sign corresponds to a soliton and the plus to an
antisoliton.  We have also
$$
(\alpha_w, q_{h-\nu})=\exp(\pi i/4 -\pi i \nu/2h) y_w^\nu,\eqno(3.43)
$$
so from eqn~(3.27) we obtain
$$
Q_{-N} = 2 (-1)^N m^N\exp\left(-(N-1){\pi i/ 4}
+(N-\nu){\pi i/2h}\right)x_w^\nu\eqno(3.44)
$$
for a soliton.

In the case $\alpha=-\alpha_b$, we have
$$
\rho=\mp\exp(\pi i/4 + \pi i/2h)\eqno(3.45)
$$
and
$$
(-\alpha_b,q_{h-\nu})=\exp(\pi i/4 + \pi i\nu/2h) y_b^\nu,\eqno(3.46)
$$
so
$$
Q_{-N} = 2 (-1)^N m^N\exp\left(-(N-1)\pi i/4 -(N-\nu)\pi i/2h\right)x_b^\nu
\eqno(3.47)
$$
for a soliton solution.

Since $N-\nu$ is equal to 0 (mod $h$), it follows that
$$
\left( \exp\bigl((N-\nu)\pi i/2h\bigr) x_w^\nu,\;
\exp\bigl(-(N-\nu)\pi i/2h\bigr) x_b^\nu\right)\eqno(3.48)
$$
is a left eigenvector of the matrix~$C$ with eigenvalue $2\cos(N\pi/h)$.
Thus we arrive finally at the result that the values of the conserved charge
$Q_{-N}$ for soliton solutions form a left eigenvector of the Cartan matrix of
$\g$, with eigenvalue $2(1-\cos N\pi/h)$.  By a suitable choice of phase all
of these values can be taken to be real.
The antisoliton solutions can be
treated similarly; these are related to the solitons by
$$
Q_{-N}({\rm antisoliton}) = (-1)^{N-1} Q_{-N}({\rm soliton}),\eqno(3.49)
$$
as one might expect.
\bigskip
{\twelvebf Conclusions}

We have shown how to calculate all conserved charges for single soliton
solutions in an affine Toda theory based on an arbitrary untwisted affine Lie
algebra $\hg$.  The values of these charges form left eigenvectors of the
Cartan matrix of the finite-dimensional Lie algebra $\g$ corresponding to
$\hg$.  Our results provide further evidence for the existence of a
unitary theory contained within an affine Toda theory with imaginary
coupling and for the duality of such a theory with an affine Toda
theory associated to the algebra $\check \g$.
It would be interesting to calculate quantum corrections~[36,37] to the results
we have obtained, although this is likely to be much more difficult than
the classical calculation we have presented.
\bigskip
{\bf Acknowledgement}

I am grateful to the UK Particle Physics and Astronomy Research Council
for financial support.

\bigskip
{{\twelvebf References}}
\parindent 15pt
\item{[1]} G. Wilson,
{\sl Ergod.\ Th.\ and Dynam.\ Sys.\ \bf1} (1981), 361.

\item{[2]} D. I. Olive and N. Turok,
{\sl Nucl.\ Phys.\ \bf B257} (1985), 277.

\item{[3]} V. G. Drinfeld and V. V. Sokolov,
{\sl J.\ Sov.\ Math.\ \bf30}
(1984), 1975--2036.

\item{[4]} H. Aratyn, L. A. Ferreira, J. F. Gomes and A. H. Zimerman,
hep-th/9308086, to appear in {\sl\IJMP\bf A}.

\item{[5]} H. W. Braden, E. Corrigan, P. E. Dorey and R. Sasaki,
proceedings of the XVIII International Conference on Differential
Geometric Methods in Theoretical Physics and Geometry, Lake Tahoe,
USA, July 1989.

\item{[6]} H. W. Braden, E. Corrigan, P. E. Dorey and R. Sasaki,
{\sl\NP\bf B338} (1990), 689--746.

\item{[7]} P. G. O. Freund, T. Klassen and E. Melzer,
{\sl\PL\bf B229} (1989), 243.

\item{[8]} T. R. Klassen and E. Melzer,
{\sl\NP\bf B338} 485--528.

\item{[9]} P. E. Dorey, unpublished.

\item{[10]} M. R. Niedermaier,
hep-th/9401078.

\item{[11]} P. Christe,
proceedings of the XVIII International Conference on Differential
Geometric Methods in Theoretical Physics and Geometry, Lake Tahoe,
USA, July 1989.

\item{[12]} P. Christe and G. Mussardo,
{\sl\NP\bf B330} (1990), 465.

\item{[13]} H. W. Braden,
{\sl J. Math.\ Phys.\ \bf A25} (1992), L15--L20.

\item{[14]} P. E. Dorey,
{\sl\NP\bf B358} (1991), 654.

\item{[15]} M. D. Freeman,
{\sl\PL\bf B261} (1991), 57.

\item{[16]} A. Fring, H. C. Liao and D. I. Olive,
{\sl\PL\bf B266} (1991), 82.

\item{[17]} T. J. Hollowood,
{\sl\NP\bf B384} (1992), 523--540.

\item{[18]} D. I. Olive, N. Turok and J. W. R. Underwood,
{\sl\NP\bf B401} (1993), 663--697.

\item{[19]} D. I. Olive, N. Turok and J. W. R. Underwood,
{\sl\NP\bf B409} (1993), 509--546.

\item{[20]} N. J. MacKay and W. A. McGhee,
{\sl\IJMP\bf A8} (1993), 2791, erratum {\sl\IJMP\bf A8} (1993), 3830.

\item{[21]} Z. Zhu and D. G. Caldi,
hep-th/9307175.

\item{[22]} C. P. Constantinidis, L. A. Ferreira, J. F. Gomes and
A. H. Zimerman,
{\sl\PL\bf B298} (1993) 88.

\item{[23]} H. Aratyn, C. P. Constantinidis, L. A. Ferreira, J. F. Gomes and
A. H. Zimerman,
presented at the VII J. A. Swieca Summer School, Particles and Fields,
Brasil, January 1993, hep-th/9304080.

\item{[24]} H. Aratyn, C. P. Constantinidis, L. A. Ferreira, J. F. Gomes and
A. H. Zimerman,
{\sl\NP\bf B406} (1993), 727.

\item{[25]} M. Niedermaier,
{\sl Comm.\ Math.\ Phys.\ \bf 160} (1994) 391.

\item{[26]} M. Niedermaier,
{\sl Lett.\ Math.\ Phys.\ \bf 30} (1994) 217.

\item{[27]} V. G. Kac,
``Infinite-dimensional Lie algebras,''
{\sl Cambridge University Press} (1990).

\item{[28]} O. Babelon and L. Bonora,
{\sl\PL\bf B244} (1990), 220--226.

\item{[29]} A. N. Leznov and M. V. Saveliev,
{\sl Lett.\ Math.\ Phys. \bf 3} (1979), 489--494.

\item{[30]} E. Date, M. Kashiwara, M. Jimbo and T. Miwa,
in Proceedings of RIMS symposium on non-linear integrable systems---classical
theory and quantum theory, Kyoto, Japan May 13--May 16, 1981, ed. by
M. Jimbo and T. Miwa (World Scientific Publishing Co., Singapore, 1983).

\item{[31]} O. Babelon and D. Bernard,
{\sl Int.\ J.\ Mod.\ Phys.\ \bf A8} (1993) 507.

\item{[32]} J. Balog, L. Feher, L. O'Raifeartaigh, P. Forgacs and A. Wipf,
{\sl Ann.\ Phys.\ \bf203} (1990), 76--136.

\item{[33]} B. Kostant,
{\sl Amer.\ J.\ Math.\ \bf81} (1959), 973--1032.

\item{[34]} H. S. M. Coxeter,
{\sl Duke.\ Math.\ Jour.\ \bf18} (1951), 765--782.

\item{[35]} A. J. Coleman,
{\sl Can.\ J.\ Math.\ \bf10} (1958), 349--356.

\item{[36]} T. Hollowood,
{\sl\PL\bf B300} (1993),73--83.

\item{[37]} G. M. T. Watts,
hep-th/9404065.

\end